\documentstyle[buckow,epsf]{article}

\def\be{\begin{equation}}
\def\ee{\end{equation}}
\def\beq{\begin{eqnarray}}
\def\eeq{\end{eqnarray}}

\def\inbar{\vrule height1.5ex width.4pt depth0pt}
\def\Cop{\relax\,\hbox{$\inbar\kern-.3em{\rm C}$}}
\def\Rop{\relax{\rm I\kern-.18em R}}
\def\Nop{\relax{\rm I\kern-.18em N}}
\def\Zop{{{\sf Z\!\!Z}}}

\def\half{{1\over 2}}

\def\ie{{\it i.e.}}
\def\H{{\cal H}}
\def\A{{\cal A}}

\def\su{\hbox{$\widehat{\rm su}$}}
\def\Vir{\hbox{\sl Vir}}

\def\beq{\begin{equation}}
\def\eeq{\end{equation}}

\begin {document}
\noindent KCL-MTH-02-02 \\
hep-th/0201113
\def\titleline{
D-branes from conformal field theory 
}
\def\authors{
Matthias R.\ Gaberdiel}
\def\addresses{
Department of Mathematics \\
King's College London \\
Strand \\
London WC2R 2LS \\
United Kingdom \\
{\tt mrg@mth.kcl.ac.uk} 
}
\def\abstracttext{
An introduction to the construction of D-branes using conformal field
theory methods is given. A number of examples are discussed in detail,
in particular the construction of all conformal D-branes for the
theory of a single free boson on a circle.} 
\large
\makefront

\section{Introduction}

D-branes in string theory can be described and analysed in essentially
two different ways. First, one can think of D-branes as being extended
objects in space-time that can wrap around certain cycles in the target
space geometry. From this point of view, D-branes are described by
geometrical data such as cohomology and K-theory
\cite{MM,WittenK,Douglas}. The second point of view, on the other
hand, is not geometrical at all: D-branes correspond to (additional)
open string sectors that can be added consistently to a given (closed) 
string theory. In this `world-sheet' approach, D-branes are described
by (boundary) conformal field theory. 

The boundary conformal field theory description is an {\it exact}
string theory description, but it is often only available at specific
points in the moduli space of target space geometries, such as
orbifold points \cite{dm,Frau1,Frau2,Sen6,BG3,DDG,DG,GabSen,GS}, Gepner
points in Calabi-Yau manifolds \cite{RS,GutSat,FSW}, {\it etc.}  On the
other hand, the geometrical approach is generically available, but it
can only be trusted whenever we are in a regime where the supergravity
approximation is good. The two approaches are therefore in some sense
complementary, and one can learn interesting features about `stringy
geometry' by comparing their results (see for example \cite{BDLR} and
the lectures of Wolfgang Lerche at this school).

In this lecture I shall attempt to give a pedagogical introduction 
to the conformal field theory approach. I shall also explain some
recent work (in collaboration with Andreas Recknagel and G\'erard
Watts \cite{GRW,GR}) that suggests that the conformal field theory
approach is tractable for larger classes of theories than had hitherto
be thought.

\section{The boundary conformal field theory approach}

Suppose we are given a conformal field theory defined on closed
Riemann surfaces, \ie\ a closed string theory. The main question we
want to address is: how can we extend this conformal field theory to a 
theory that is also defined on world-sheets with boundary. More
precisely we want to ask which boundary conditions can be
imposed at the various boundaries. From a string theory point of view,
this is the question of  which open strings can be consistently added
to a given closed string theory. 

In some sense, this problem is rather similar to a familiar
construction in (closed) conformal field theory. Suppose we are given
the theory defined on the sphere. We can then ask whether this theory 
determines already (uniquely) the theory on arbitrary Riemann
surfaces. The answer is well known \cite{Vafa,Sonoda,MooreSei}: the
theory on the sphere determines uniquely the theory on an arbitrary
closed Riemann surface (if it exists), but it does not guarantee that
it is consistent. Indeed, there is one additional consistency
condition that arises at genus $1$ (and that does not follow from the
consistency of the theory on the sphere), namely that the correlation 
functions on the torus transform under the action of the modular group
SL$(2,\Zop)$.\footnote{For example, the theory of a single NS fermion
is consistent on the sphere but does not satisfy the modular
consistency condition.} If this consistency condition is satisfied,
the theory is consistent on all Riemann surfaces \cite{MooreSei}.   

The analogous result for the construction of the theory on surfaces
with boundaries is not really known. For a given theory defined on the 
sphere, the complete list of `sewing relations' that have to be
satisfied by each boundary condition is known
\cite{lew,carlew}. However it is not clear for which classes of
theories solutions to these sewing relations can be found, and if so,
how many. Based on the examples that have been understood
\cite{PSS1,Ingo1,BPPZ,Ingo2,FFFS1,FFFS2} it appears that modular
invariance may again be sufficient to guarantee that a `complete' set
of boundary conditions can be constructed. In fact, there are
striking similarities between the classification of modular invariant
partition functions and that of the so-called NIM-reps (non-negative
integer matrix representations of the fusion algebra) that appear
naturally in the construction of the boundary states
\cite{BPPZ,Gannon}. On the other hand, it seems that there are more
NIM-reps than (consistent) conformal field theories that can be
defined on the torus, and at least some of the additional NIM-reps
seem to be naturally related to consistent conformal field theories
that are only defined on the sphere (but not on the torus). 

The basic reason why the theory on the sphere determines already the
theory on all Riemann surfaces can be schematically understood as
follows. Since we are dealing with a local conformal field theory, the
operator product expansion of any two operators is the same, 
irrespective of the surrounding surface. The operator product
expansion (and thus the `local structure' of the theory) is therefore
already determined by the theory on the sphere. 

For the problem we are actually interested in, namely the extension of
the theory on the sphere, say, to surfaces with boundary, a similar
consideration applies. As we have just explained, given the theory on 
the sphere we can deduce the operator product expansion of the fields 
$\phi_a$, which we can write schematically as 
\be\label{ope}
\phi_a \, \phi_b \sim \sum_c C_{ab}^c \phi_c \,.
\ee
Here $C_{ab}^c$ are the structure constants of the theory on the
sphere, and we have suppressed the dependence of the fields on the
coordinates on the sphere. We can think of the operator product
expansion as defining an `algebra of fields'.\footnote{Because of the
dependence on the coordinates, this is not really an algebra, but
rather (a slight generalisation of) what is usually called a 
{\it vertex operator algebra}.} The boundary conditions we are
interested in have to respect this algebra, and they must therefore
define an `algebra homomorphism' 
\be\label{map}
\left( \hbox{`algebra of fields'} \right) \longrightarrow \Cop \,.
\ee
Every element of the space of states of the theory on the sphere $\H$ 
defines a map of the form (\ref{map}), and in fact every such map
arises from a suitable (infinite) linear combination of such states.  
Thus we can describe each boundary condition by a `coherent' 
{\it boundary state} in $\H$\footnote{As we shall see momentarily, the
boundary states are necessarily coherent states, \ie\ they do not lie
in the Fock space of finite energy states.}, and for the boundary 
condition labelled by $\alpha$ we denote the corresponding boundary
state by $|\!|\alpha\rangle\!\rangle$. Given this boundary state, the   
amplitudes of the fields in the presence of the boundary with boundary
condition $\alpha$ are then simply given by the (closed string)
expression   
\be
\langle \phi_1 \, \phi_2\, \phi_3 \rangle_{\alpha} =
\langle  \phi_1 \, \phi_2\, \phi_3 \, |\!|\alpha\rangle\!\rangle\,.
\ee
\smallskip

\noindent Not every linear map of the form (\ref{map}) actually
defines a boundary state. (Indeed, it follows from the above
discussion, that there exists for example a coherent state for each
higher genus Riemann surface.) The coherent states that describe
boundary conditions are characterised by the property that the left-
and right-moving fields corresponding to unbroken symmetries are
related to one another at the boundary. If we take the boundary to be
along the real axis, the relevant condition is that
\be
S(z) = \rho\Bigl( \bar S (\bar{z}) \Bigr)\qquad 
\hbox{for $z\in\Rop$,} 
\ee
where $S$ and $\bar{S}$ are generators of the symmetry that is
preserved by the boundary, and $\rho$ denotes an automorphism of the
algebra of fields that leaves the stress-energy tensor invariant. The
fields $S(z)$ and $\bar S(\bar{z})$ have an expansion in terms of
modes as   
\be\label{modes}
S(z) = \sum_{n\in\Zop} S_n z^{-n-h} \,, \qquad
\bar{S}(\bar{z}) = \sum_{n\in\Zop} \bar{S}_n \bar{z}^{-n-h} \,,
\ee
where $h$ is the conformal weight of $S$ (and $\bar{S}$). In the
description of the boundary condition in terms of a boundary state,
the boundary is taken to be the (unit) circle around the origin. In
order to express the above condition in terms of a condition involving
the boundary state, we apply the conformal transformations 
$\zeta(z)=e^{- 2 \pi \imath z}$ and 
$\bar\zeta(\bar{z})=e^{+ 2 \pi \imath \bar{z}}$ to obtain 
\be
(-2\pi\imath\, \zeta)^h\, S(\zeta) = 
(2\pi\imath\, \bar\zeta)^h\, \rho\Bigl( \bar S(\bar\zeta) \Bigr)\qquad
\hbox{for $|\zeta|=1$,}
\ee
where we have used that a (primary) conformal field transforms as 
\be
S(z) \mapsto {\zeta'(z)}^h S(\zeta(z))\,, 
\ee
and similarly for $\bar{S}$. Using (\ref{modes}) we thus find that a
boundary state $|\!|\alpha\rangle\!\rangle$ that preserves the
symmetry described by $S$ has to satisfy
\be\label{gluing0}
\left( \sum_{n\in\Zop} S_n\, \zeta^{-n} 
- (-1)^h \sum_{n\in\Zop} \rho\Bigl(\bar{S}_n\Bigr)\, 
\bar\zeta^{-n} \right) 
|\!|\alpha\rangle\!\rangle\ = 0 \qquad \hbox{for $|\zeta|=1$.}
\ee
Since this has to hold for all $\zeta$ with $|\zeta|=1$,
(\ref{gluing0}) implies the so-called `gluing condition'
\be\label{gluing}
\left( S_n - (-1)^{h} \rho\Bigl(\bar{S}_{-n}\Bigr) \right)
|\!|\alpha\rangle\!\rangle\ = 0 
\qquad \hbox{for all $n\in\Zop$.}
\ee
The gluing condition implies in particular that 
$|\!|\alpha\rangle\!\rangle$ must be a coherent state since no vector
in the Fock space of finite energy states can satisfy (\ref{gluing})
for all $n\in\Zop$.

There is only one symmetry that every boundary condition has to
preserve. This is the `conformal' symmetry that guarantees that the 
resulting field theory is again conformal. In terms of (\ref{gluing})
it corresponds to the gluing condition 
\be\label{conformal}
\left( L_n - \bar{L}_{-n} \right) |\!|\alpha\rangle\!\rangle\ = 0
\qquad \hbox{for all $n\in\Zop$,}
\ee
where $L_n$ and $\bar{L}_n$ are the modes of the left- and
right-moving stress energy tensor of conformal weight 
$h_L=h_{\bar L}=2$. If we impose other gluing conditions beyond
(\ref{conformal}) we are restricting our attention to special boundary 
conditions. However, from an abstract point of view, one would expect
that all conformal boundary conditions will play a r\^ole.

The more symmetries we require the boundary condition to preserve, the
fewer boundary conditions exist, and the more tractable the problem 
becomes. The situation is particularly simple if the closed theory
is a `rational' theory with respect to the preserved symmetry
algebra: let us assume we are interested in boundary conditions that
respect the symmetry algebra $\A$ (where we take, for simplicity, 
$\rho=\hbox{id}$). In order to determine the relevant boundary
conditions we decompose the space of states of the closed string
theory $\H$ in terms of representations of $\A\otimes\bar\A$ as    
\be\label{decomposition}
\H = \bigoplus_{i,j} N_{ij}\, \H_i \otimes \bar\H_j \,,
\ee
where the sum runs over the set of irreducible representations of $\A$
and $\bar\A\cong \A$, and $N_{ij}$ describes the multiplicity with
which the irreducible representation $\H_i \otimes \bar\H_j$ of
$\A\otimes\bar\A$ appears in $\H$. The theory is called `rational with
respect to $\A$' if $\A$ only possesses finitely many irreducible
representations. In this case, the sum in (\ref{decomposition}) is
finite.  

Since the modes that appear in the gluing condition (\ref{gluing}) map
each $\H_i\otimes\bar\H_j$ into itself, we can solve the gluing
constraint separately for each summand in (\ref{decomposition}). We
can find a non-trivial solution provided that $\H_i$ is the conjugate
representation of $\bar\H_j$. If this is the case, there is (up to
normalisation) only one coherent state that satisfies 
(\ref{gluing});\footnote{We are assuming here, for ease of notation,
that the multiplicities $N_{ij}$ are all either zero or one; the 
modifications for the general case are obvious.} this state is called
the {\it Ishibashi state} \cite{Ishibashi} and it is denoted by  
\be\label{Ishibashi}
|i\rangle\!\rangle \in \H_i\otimes\bar\H_i\,, \qquad 
\left( S_n - (-1)^{h_S} \rho(\bar{S}_{-n}) \right) 
|i\rangle\!\rangle\ =0 \qquad \hbox{for all $n\in\Zop$ and $S\in\A$.}
\ee
If the theory is rational then there are in particular only finitely
many Ishibashi states. 

Since every boundary state satisfies the gluing condition
(\ref{gluing}) it must be a linear combination of the Ishibashi
states. We can therefore write every boundary state as 
\be\label{ansatz}
|\!|\alpha\rangle\!\rangle\ = 
\sum_i B_\alpha^i \, |i\rangle\!\rangle \,,
\ee
where $B_\alpha^i$ are some constants that characterise the boundary
condition. These constants are constrained by two classes of
conditions: 
\begin{list}{(\roman{enumi})}{\usecounter{enumi}}
\item The Cardy condition \cite{Cardy}.
\item The so-called `sewing relations' that were first derived in
\cite{carlew,lew}. 
\end{list}
The Cardy condition (i) comes about as follows. Let us consider the
(open string) partition function 
\be\label{open}
Z_{\alpha\beta}(\tilde{q}) = \hbox{Tr}_{\H_{\alpha\beta}}
e^{-2\pi T H_{o}} = \sum_i n^i_{\alpha\beta}
\chi_i(\tilde{q}) 
\ee
of the open string with boundary conditions $\alpha$ and $\beta$ at
the two ends. Here $\H_{\alpha\beta}$ is the corresponding space of
open string states, and $H_{o}$ the open string Hamilton
operator (that equals $H_{o}=L_0-{c\over 24}$). In writing the second
equation in (\ref{open}) we have used that the boundary conditions
preserve $\A$, and therefore that we can decompose $\H_{\alpha\beta}$
with respect to $\A$ as 
\be
\H_{\alpha\beta} = \bigoplus_i n^i_{\alpha\beta} \H_i \,,
\ee
where each $\H_i$ is an irreducible representation of $\A$. The
numbers $n^i_{\alpha\beta}$ describe the multiplicity with which
$\H_i$ appears in $\H_{\alpha\beta}$, and they are therefore
non-negative integers. (In fact, the numbers $n^i_{\alpha\beta}$ are
precisely the entries of the NIM-reps we mentioned before.) We have
furthermore used the usual short hand notation for the character of a
representation,  
\be\label{opencalc}
\chi_i(\tilde{q}) 
= \hbox{Tr}_{\H_i} \left(e^{-2\pi T 
\left(L_0 - {c\over 24}\right)}\right)\,,
\qquad \tilde{q} = e^{-2\pi T} \,.
\ee
In terms of the boundary states we introduced before (\ie\ from the
closed string point of view) this amplitude is simply the overlap 
\begin{equation}
\langle\!\langle \alpha |\!|\, e^{-2\pi L H_{cl}}\,
 |\!| \beta \rangle\!\rangle 
 =  \sum_i \left(B^i_\alpha\right)^\ast \, B^i_\beta \; \chi_i(q)
\,. \label{calcone}
\end{equation}
Here $H_{cl}$ is the closed string Hamiltonian, and we have used 
(\ref{ansatz}) to write the boundary states in terms of the Ishibashi
states. We have furthermore used that 
\be
\langle\!\langle i |\, e^{-2\pi L H_{cl}}\, 
| j \rangle\!\rangle =  \delta_{ij} \chi_i(q) \,,
\ee
where $\chi_i(q)$ is again the character of the representation $\H_i$
that is now evaluated at $q$ with $q = e^{-2 \pi L}$ rather than
$\tilde{q}$. In relating the open- and closed point of view, \ie\ in
relating (\ref{open}) and (\ref{calcone}), we have to exchange
what we think of as being the time- and space-coordinate on the
world-sheet; we therefore have to identify $T=1/L$. If we write 
\be
q = e^{2\pi i \tau}\,,\qquad \tau=iL\,,
\ee
then $\tilde{q}$ is simply given as 
\be
\tilde{q} = e^{- {2 \pi i \over \tau}} \,.
\ee
Thus $q$ and $\tilde{q}$ are related by the standard modular
$S$-transformation that maps $\tau\mapsto-1/\tau$. At least for
rational conformal field theories (and in fact under certain slightly
weaker conditions) the characters of the irreducible representations
transform into one another as   
\be\label{modular} 
\chi_i(q) = \sum_j S_{ji}\, \chi_j(\tilde{q}) \,,
\ee
where $S_{ij}$ is the symmetric and unitary matrix representing the 
$S$-transformation of the modular group SL$(2,\Zop)$. Inserting
(\ref{modular}) into (\ref{calcone}) we therefore find that  
\be
Z_{\alpha\beta}(\tilde{q}) = 
\sum_{i,j} \left(B^i_\alpha\right)^\ast \, B^i_\beta \, S_{ji} \,
\chi_j(\tilde{q}) \,.
\ee
Comparing with (\ref{open}), and assuming that the characters of 
the irreducible representations are linearly independent, it therefore 
follows that 
\be\label{NIM}
n^j_{\alpha\beta} = \sum_{i} \left(B^i_\alpha\right)^\ast \, B^i_\beta
\, S_{ji}\,.
\ee
This is a very restrictive condition that is often (in particular, if
the theory is rational and there are only finitely many irreducible
representations) fairly accessible. 
\smallskip

Before proceeding we should note that the set of solutions to Cardy's
condition form (the positive cone of) a lattice: suppose that the set 
\be
M = \left\{ |\!| \alpha_1 \rangle\!\rangle\,, \ldots \,, 
|\!| \alpha_n \rangle\!\rangle\right\}
\ee
satisfies Cardy's condition, \ie\ the overlap between any two elements
of $M$ leads to non-negative integer numbers
$n^i_{\alpha_j,\alpha_k}$, then so does the set 
\be
M' = \left\{ |\!| \alpha_1 \rangle\!\rangle\,, \ldots \,, 
|\!| \alpha_n \rangle\!\rangle \,, 
\sum_{l=1}^{n} m_l |\!| \alpha_l \rangle\!\rangle \right\}\,,
\ee
provided that $m_l\in\Nop_0$ for $l=1,\ldots, n$. This is simply a 
consequence of the fact that sums of products of non-negative integers
are non-negative integers. What we therefore want to find are the 
{\it fundamental} boundary conditions that generate all other boundary
conditions upon taking positive integer linear combinations as
above. 

In general, rather little is known about how to characterise these
fundamental D-branes intrinsically. Part of the problem is due to the
fact that the Cardy condition by itself does not have a unique
solution. In order to illustrate this fact let us consider one of the
simplest conformal field theories, the so-called Ising model. This
theory is the minimal Virasoro model with $c={1\over 2}$ for which 
\be
\H= \left(\H_0\otimes\bar\H_0\right) \bigoplus 
\left(\H_{1\over 2}\otimes\bar\H_{1\over 2}\right)
\bigoplus \left(\H_{1\over 16} \otimes \bar\H_{1\over 16}\right)\,.
\ee
Here $\H_h$ is the irreducible representation of the Virasoro algebra
for which the highest weight state has $L_0$ eigenvalue equal to
$h$. In the basis corresponding to $\H_0,\H_{1\over 2},\H_{1\over 16}$, 
the $S$-matrix is given as 
\be
S = {1\over 2} \left( \begin{array}{ccc}
1 & 1 & \sqrt{2} \\
1 & 1 & - \sqrt{2} \\
\sqrt{2} & - \sqrt{2} & 0 
\end{array}
\right)\,.
\ee
One set of (fundamental) boundary states can be constructed following
the general procedure of Cardy \cite{Cardy} as follows. For each
irreducible representation $\H_i$ we define the boundary state 
$|\!|\alpha_i \rangle\!\rangle$ in terms of the Ishibashi states by
\be
|\!| \alpha_j \rangle\!\rangle = 
\sum_l {S_{jl}\over \sqrt{S_{0l}}} \, | l \rangle\!\rangle\,.
\ee
For these boundary states, the corresponding integers
$n^i_{\alpha_j\alpha_k}$ are then simply the fusion matrices, since
(\ref{NIM}) becomes in this case 
\be
n^i_{\alpha_j\alpha_k} = \sum_l {S_{jl}^\ast\over \sqrt{S_{0l}}} 
{S_{kl}\over \sqrt{S_{0l}}} S_{il} = 
\sum_l {S_{il} S_{kl} S_{jl}^\ast \over S_{0l}}\,,
\ee
which equals the fusion matrix $N^j_{ik}$ by Verlinde's formula
\cite{Verlinde}. 

\noindent Applying the general formula to the case at hand, the three
Cardy boundary conditions are then
\begin{eqnarray}
|\!| \alpha_0 \rangle\!\rangle & = & 
{1\over\sqrt{2}}\, | 0 \rangle\!\rangle 
+ {1\over\sqrt{2}}\, | {1\over 2} \rangle\!\rangle 
+ {1\over 2^{1\over 4}}\, | {1\over 16} \rangle\!\rangle \,, \nonumber
\\
|\!| \alpha_{1\over 2} \rangle\!\rangle & = & 
{1\over\sqrt{2}}\, | 0 \rangle\!\rangle 
+ {1\over\sqrt{2}}\, | {1\over 2} \rangle\!\rangle 
- {1\over 2^{1\over 4}}\, | {1\over 16} \rangle\!\rangle \,,
\label{cardystates} \\
|\!| \alpha_{1\over 16} \rangle\!\rangle & = & 
| 0 \rangle\!\rangle - | {1\over 2} \rangle\!\rangle \,.\nonumber 
\end{eqnarray}
However, this does not define the only solution to Cardy's
conditions. Indeed, another set of `fundamental' boundary states is
given by the three boundary states
\begin{eqnarray}
|\!| 1 \rangle\!\rangle & = & | 0 \rangle\!\rangle 
+ | {1\over 2} \rangle\!\rangle 
+ 2^{1\over 4}\, | {1\over 16} \rangle\!\rangle
\,, \nonumber
\\
|\!| 2 \rangle\!\rangle & = & | 0 \rangle\!\rangle 
+ | {1\over 2} \rangle\!\rangle \,, \label{crazy}
\\
|\!| 3 \rangle\!\rangle & = & \sqrt{2} \, | 0 \rangle\!\rangle 
- \sqrt{2}\, | {1\over 2} \rangle\!\rangle \,.\nonumber 
\end{eqnarray}
It is not difficult to check that these three boundary states also
satisfy Cardy's condition, \ie\ that the relative overlaps give rise
to non-negative integers (\ref{NIM}). However, these two sets of
boundary states are mutually incompatible since, for example, the
overlap 
\be
\langle\!\langle \alpha_0 |\!|\, e^{-2\pi L H_{cl}}\,
 |\!| 1 \rangle\!\rangle = \sqrt{2} \chi_0(\tilde{q}) \,.
\ee
While both of these sets of boundary states satisfy Cardy's condition,
it is believed that only (\ref{cardystates}) actually defines
consistent boundary states, and that the other consistency conditions
(some of which we are about to discuss) will rule out
(\ref{crazy}). However, this example demonstrates that Cardy's
consistency condition alone does not allow one to discard
(\ref{crazy}).  
\bigskip

\noindent Suppose now that we have found a solution to Cardy's
condition that also satisfies all other consistency conditions. We
would like to find a good criterion that allows us to find the
fundamental boundary states out of which all other boundary states can
be obtained as positive integer linear combinations. As we mentioned
before, little is known about this in general, but it is believed that
the fundamental boundary states are characterised by the condition
that the self-overlap of each such boundary state 
$|\!| \alpha\rangle\!\rangle$ contains the vacuum representation in
the open string channel with multiplicity one, \ie\    
\be\label{condition}
\langle\!\langle \alpha |\!|\, e^{-2\pi L H_{cl}}\,
 |\!| \alpha \rangle\!\rangle = \chi_0(\tilde{q}) + \cdots \,.
\ee
With this criterion any non-trivial integer linear combination of
fundamental boundary states is then not fundamental. Incidentally,
according to this criterion the three Cardy states are fundamental, 
but for example $|\!| 1 \rangle\!\rangle$ is not. In the following we
shall call a family of boundary states fundamental if
(\ref{condition}) is satisfied for all its members. 
\bigskip

For D-branes that are fundamental in this sense, one of the sewing
relations simplifies considerably, and actually gives rise to a
powerful constraint (see \cite{GRW} for a more detailed
derivation). This constraint arises from considering a two-point 
function of primary fields in the presence of such a boundary
condition, 
\newpage

\begin{figure}[h]
\hspace*{2cm}
\epsfxsize=4cm
\epsfbox{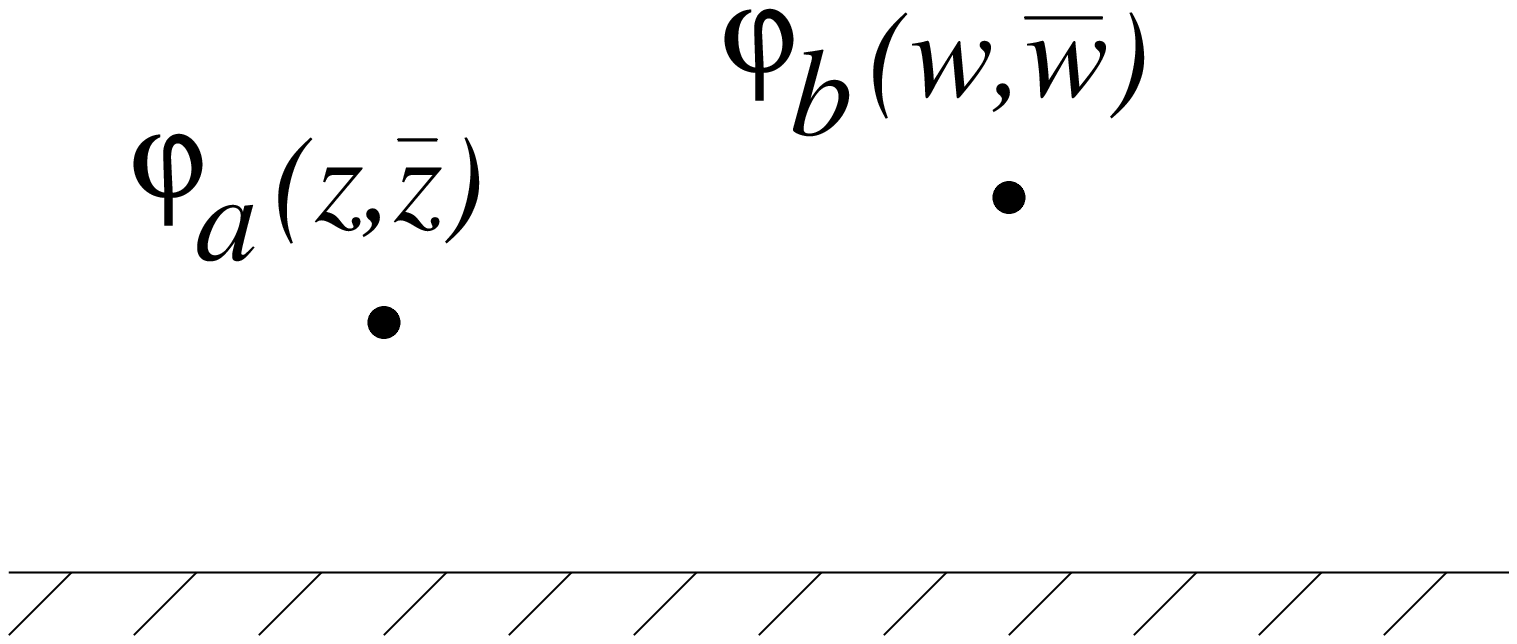}
\vspace*{-1cm}
\be\hspace*{5cm}\label{tpfa}
F_{ab}(z,\bar z,w,\bar w)
= \Big\langle
  \; \varphi_a(z,\bar z) \; \varphi_b(w,\bar w) \;
  \Big\rangle
\,.
\ee
\end{figure}

\noindent The gluing conditions for the energy-momentum tensor imply
that (\ref{tpfa}) can be described in terms of four-point chiral
blocks where we insert chiral vertex operators of weight 
$h_a, \bar h_a, h_b$ and $\bar h_b$ at $z,\bar z, w$ and $\bar w$,
respectively. This four-point function can then be factorised in two
different ways, leading to two different representations of the
correlation function, as shown below. In the first picture one
considers the limit in which the two fields approach the boundary
separately; in the second picture on the other hand, the two fields
come close together away from the boundary, and we can thus use the
operator product expansion (\ref{ope}) in order to express the product
of these two fields in terms of a sum of single fields:
\begin{figure}[h]
\vspace*{0cm}\hspace*{0.5cm}
\epsfxsize=4cm
\epsfbox{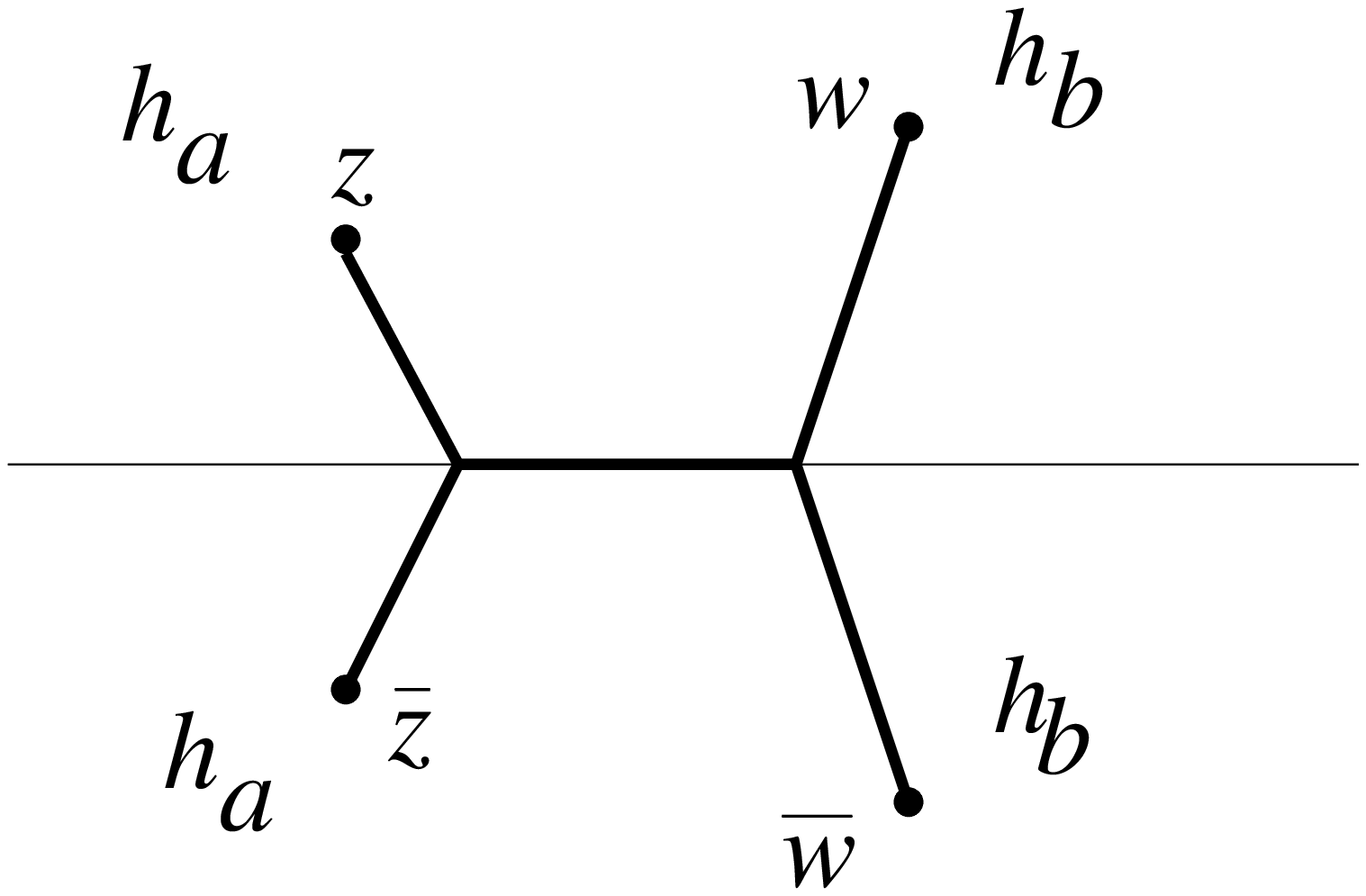}
\vspace*{-2.1cm}
\be\hspace*{2.2cm}\label{cblocksa}
\sim B_\alpha^a B_\alpha^b \,
|z-\bar z|^{2h_b - 2h_a} \,
|z - \bar w|^{-4 h_b} \,
f^{1}\!\left[{\scriptstyle{a\, b}\atop \scriptstyle{a\, b}}\right] 
(\eta) \,.
\ee
\end{figure}
\begin{figure}[h]
\vspace*{1cm}\hspace*{0.5cm}
\epsfxsize=4cm
\epsfbox{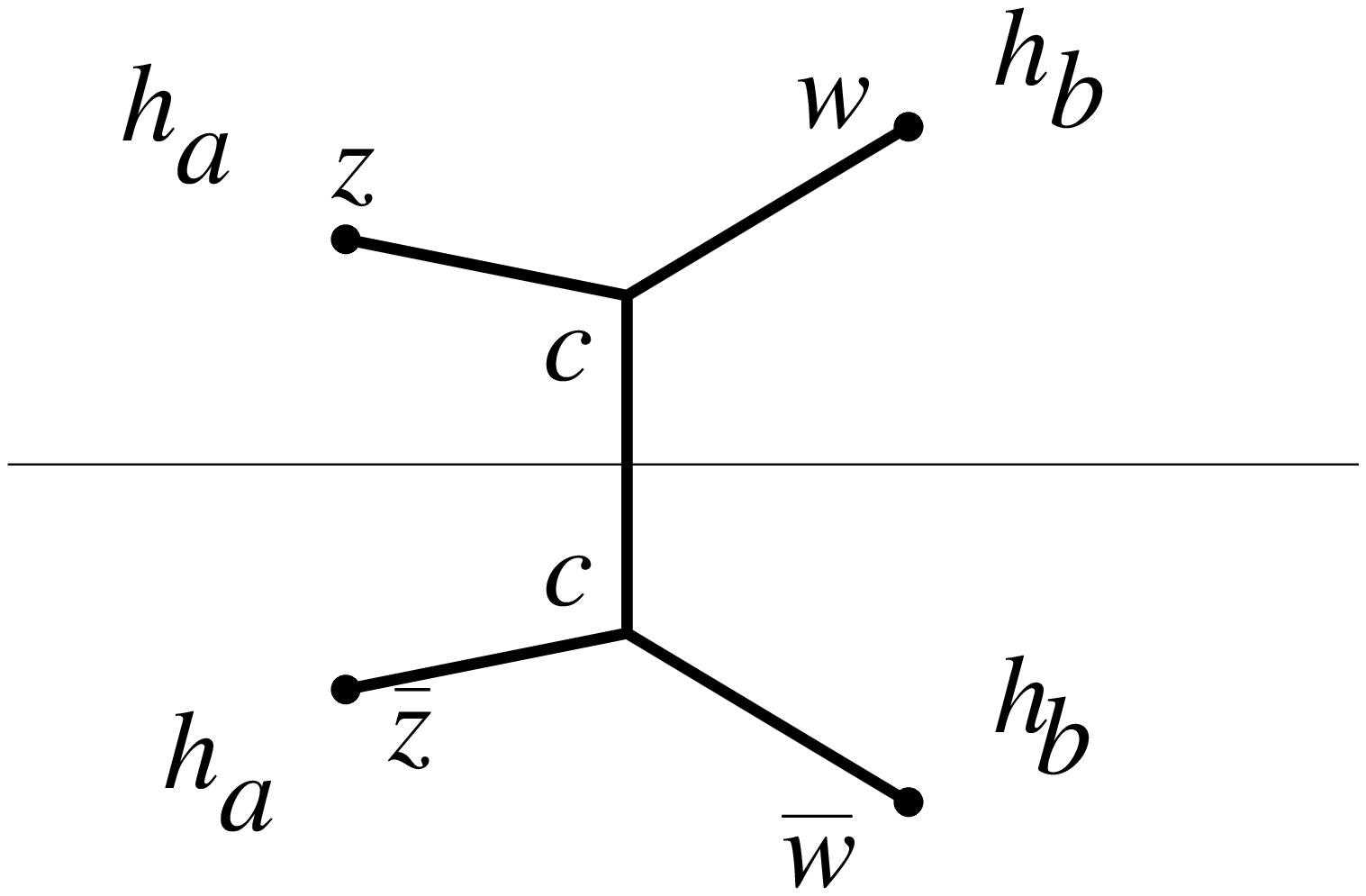}
\vspace*{-2.1cm}
\be\hspace*{4.5cm}\label{cblocksb}
\sim
\sum_{c} \,
C_{ab}{}^c \,
B_\alpha^c \,
|z-\bar z|^{2h_b - 2h_a} \,
|z - \bar w|^{-4 h_b} \,
f^{c}\!\left[{\scriptstyle{b\, b}\atop \scriptstyle{a\, a}}\right] 
(1-\eta) \,.
\ee
\end{figure}
\bigskip

\noindent In writing down these equations we have specialised to the
case where $\varphi_a$ and $\varphi_b$ are self-conjugate fields for
which $h_a=\bar h_a$ and $h_b=\bar h_b$. The $f^1$ and $f^c$ 
denote the different chiral four-point blocks, and $\eta$ is the
cross-ratio $\eta = |(z-w)/(z-\bar w)|^2$ which is real with 
$0\leq\eta\leq 1$. In both equations we have only considered the
leading behaviour as $\eta\rightarrow 1$, \ie\ we have only taken 
into consideration the contribution of the vacuum state in the open
string channel. (This is where we have used the assumption that the 
boundary condition in question is fundamental to deduce that there is 
only one such state.)

\noindent The two sets of chiral blocks are related by the so-called
fusing matrices 
\be\label{fusmat}
f^{c}\!\left[{\scriptstyle{b\, b}\atop \scriptstyle{a\, a}}\right] 
(1-\eta) = 
F_{c1}\!\left[{\scriptstyle{b\, b}\atop \scriptstyle{a\, a}}\right] \, 
f^{1}\!\left[{\scriptstyle{a\, b}\atop \scriptstyle{a\, b}}\right] 
(\eta)\,.
\ee
Substituting (\ref{fusmat}) in (\ref{cblocksb}) and comparing with 
(\ref{cblocksa}), we then obtain the sewing relation
\be
\label{ca}
B_\alpha^a \, B_\alpha^b \,
=
\sum_{c} \,
C_{ab}{}^c \,
F_{c1}\!\left[{\scriptstyle{b\, b}\atop \scriptstyle{a\, a}}\right]\,
B_\alpha^c\,.
\ee
This condition is known as the `factorisation constraint'
\cite{carlew,lew}, the `cluster condition' \cite{RS1} or the 
`classifying algebra' \cite{FS}. In many cases it is, however, rather
difficult to check since the structure constants $C$ and $F$ on the
right hand side are often not explicitly known. However, there are 
a few examples where one can actually determine the structure of
(\ref{ca}) explicitly, and use this to classify all possible
fundamental D-branes of a theory. We shall come back to this
point later.

\section{Some illustrative examples}

In this section we want to give some examples that will illustrate the
structure we have described above. 

\subsection{A free boson on a circle of radius $R$}

This theory is generated by the modes of the field $X(z,\bar{z})$ for 
which we identify $X$ with $X+2\pi R$. The left- and right-moving modes
are denoted by $\alpha_l$ and $\bar\alpha_l$, respectively, and they
satisfy the commutation relations
\begin{eqnarray}
{}[\alpha_l,\alpha_k] & = & l\, \delta_{l,-k} \nonumber \\
{}[\alpha_l,\bar\alpha_k] & = & 0 \nonumber \\
{}[\bar\alpha_l,\bar\alpha_k] & = & l\, \delta_{l,-k}\,. \nonumber
\end{eqnarray}
The space of states of the theory is 
\be
\H = \bigoplus_{m,n} \H_{(m,n)} \,,
\ee
where $\H_{(m,n)}$ consists of the states that are generated by the
action of the negative modes $\alpha_{-l}$ and $\bar\alpha_{-l}$ with
$l>0$ from a ground-state $|(p_L,p_R)\rangle$ for which 
\be
\alpha_0\, |(p_L,p_R)\rangle = p_L \,|(p_L,p_R)\rangle \qquad
\bar\alpha_0 \, |(p_L,p_R)\rangle = p_R  \, |(p_L,p_R)\rangle\,,
\ee
with
\be
(p_L,p_R) = \left( {m\over 2R} + n R, {m\over 2R} - n R \right) \,.
\ee
The branes that respect the U(1) symmetry corresponding to $\alpha_l$
and $\bar\alpha_l$ are either {\it Neumann} (N) or {\it Dirichlet} (D) 
branes. They are characterised by the gluing conditions
\begin{eqnarray}
\left( \alpha_l + \bar\alpha_{-l} \right) |{\rm N}\rangle\!\rangle 
& = & 0 \qquad l\in\Zop \label{Neuglu} \,, \\
\left( \alpha_l - \bar\alpha_{-l} \right) |{\rm D}\rangle\!\rangle
& = & 0 \qquad \label{Dirglu} l \in \Zop\,.
\end{eqnarray}
Since $\alpha_l$ are the modes of a field of conformal weight one, it
follows from (\ref{gluing}) that the Neumann gluing condition
corresponds to $\rho=\hbox{id}$, while the automorphism for the
Dirichlet gluing condition is $\rho(\alpha_l)=-\alpha_l$. This
automorphism leaves the stress energy tensor invariant since the
Virasoro generators are bilinear (normal ordered) products of the
$\alpha_l$ modes. It also follows from this fact that both gluing
conditions actually imply the conformal gluing condition
(\ref{conformal}). 

The condition (\ref{Neuglu}) with $l=0$ implies that a Neumann
Ishibashi state can only be constructed in $\H_{(m,n)}$ provided that 
$p_L=-p_R$. In terms of our previous discussion this is simply the
statement that the left- and right representations of the preserved
symmetry algebra must be conjugate representations. At a generic
radius $R$, $p_L=-p_R$ can only be satisfied if $m=0$, and thus we
have a Neumann Ishibashi state for each $n\in\Zop$,  
\be
\vert \, (nR,-nR)\, \rangle\!\rangle^{\rm N} \in \H_{(0,n)} \,.
\ee
Similarly, a Dirichlet Ishibashi state can only be constructed in
$\H_{(m,n)}$ provided that $p_L=p_R$; at a generic radius we
therefore only have the Dirichlet Ishibashi states
\be
\vert \,  ({m\over 2R},{m\over 2R}) \, \rangle\!\rangle^{\rm D} 
\in \H_{(m,0)} \,,
\ee
where $m\in\Zop$. In this simple example, one can actually give a
closed formula for these Ishibashi states; they are simply given as
\begin{eqnarray}
\vert \, (nR,-nR)\, \rangle\!\rangle^{\rm N} & = & 
\exp\left( \sum_{l=1}^{\infty} -{1\over l}\, \alpha_{-l} \bar\alpha_{-l}
\right) \left|(nR,-nR)\right\rangle \nonumber \\
\vert \, ({m\over 2R},{m\over 2R})\, \rangle\!\rangle^{\rm D} & = & 
\exp\left( \sum_{l=1}^{\infty} {1\over l}\, \alpha_{-l} \bar\alpha_{-l}
\right) \left|\left({m\over 2R},{m\over 2R}\right)\right\rangle \,.
\end{eqnarray}
It is easy to see that these states satisfy indeed (\ref{Neuglu}) and
(\ref{Dirglu}), respectively.

The actual D-branes (that satisfy Cardy's condition) are given as
linear combinations of these Ishibashi states. In the present case,
the relevant expressions are 
\be\label{Neubrane}
|\!|w \rangle\!\rangle =  R^\half\, \sum_{n\in\Zop} e^{i w n R} 
\vert \, (nR,-nR)\, \rangle\!\rangle^{\rm N}\,,
\ee
which describes a Neumann brane with Wilson line $w$, and
\be\label{Dirbrane}
|\!|a \rangle\!\rangle =  {1\over (2R)^\half}
\sum_{m\in\Zop} e^{i {m a \over R}}
\vert \, ({m\over 2R},{m\over 2R})\, \rangle\!\rangle^{\rm D}\,,
\ee
which corresponds to a Dirichlet brane at the position $a$. Given the
explicit form of the Ishibashi states, it is now straightforward to work
out the closed string tree diagram, \ie\ the overlap (\ref{calcone}),
and to check that the corresponding open string loop diagram, \ie\ the 
partition function (\ref{open}), satisfies the Cardy condition. All of
the branes above are fundamental. 

For both Neumann and Dirichlet branes one can also analyse the
corresponding factorisation constraint. In both cases, the relevant
classifying algebra simplifies considerably since the combination of
$C$ and $F$ that appears in (\ref{ca}) is essentially trivial. More
precisely, if we write the boundary states as 
\begin{eqnarray}
|\!|B^N \rangle\!\rangle & = & R^\half\, \sum_{n\in\Zop} \hat{B}^N_n \;
\vert \, (nR,-nR)\, \rangle\!\rangle^{\rm N} \nonumber \\
|\!|B^D \rangle\!\rangle & = & {1\over (2R)^\half} \sum_{m\in\Zop} 
\hat{B}^D_m\; \vert \, ({m\over 2R},{m\over 2R})\, 
\rangle\!\rangle^{\rm D}\,,
\end{eqnarray}
the factorisation constraint simply becomes
\begin{eqnarray}
\hat{B}^N_{n_1} \cdot \hat{B}^N_{n_2} & = & 
\hat{B}^N_{n_1+n_2} \nonumber \\
\hat{B}^D_{m_1} \cdot \hat{B}^D_{m_2} & = & 
\hat{B}^D_{m_1+m_2} \,.
\end{eqnarray}
The most general fundamental U(1)-preserving Neumann and Dirichlet
branes are thus given by 
\be
\hat{B}^N_{n} = e^{iw n R} \qquad 
\hat{B}^D_{m} = e^{i {m a \over R}}\,,
\ee
and therefore correspond to the branes given above. Strictly speaking,
one could also choose $w$ and $a$ to be arbitrary complex (rather than 
real) numbers. While the resulting branes seem to be consistent from a
conformal field theory point of view, they  have complex couplings to
some of the space-time fields, and are therefore presumably unphysical.
\bigskip

The boundary conditions that we have analysed so far are rather
special in that they preserve the full U(1)-symmetry rather than just
the conformal symmetry. From an abstract point of view, it is not
clear why these special branes should account for all the relevant
branes of the theory, and it is therefore important to understand what
other (conformal) branes exist. In order to answer this question let
us first consider the case where the radius of the circle is at the
self-dual point, $R={1\over \sqrt{2}}$. Then the theory is the same as
the WZW model corresponding to SU(2) at level $k=1$.

\subsection{The conformal branes for SU(2) level one}

Let us begin by recalling some properties of this theory on the
sphere. The theory has left- and right-moving currents $J^a(z)$ and 
$\bar J^a(\bar z)$ whose modes $J^a_n$ and $\bar J^a_n$ define two
commuting copies of the affine algebra $\widehat{\rm su}(2)_1$. The  
space of states of the theory on the sphere $\H_{\rm sphere}$ can
therefore be decomposed into highest weight representations of these
two algebras.  

There are two highest weight representations of the affine algebra
$\su(2)$ at level $1$ that actually define representations of the
conformal field theory (or the corresponding vertex operator algebra):
the vacuum representation $\H_0$ that is generated from a state
transforming in the singlet representations of the horizontal su(2)
algebra (spanned by the zero modes $J^a_0$), and the representation
$\H_{\half}$ for which the highest weight states transform in the
doublet representation of the horizontal su(2) algebra. The free boson
theory at the self-dual radius we are interested in, has a space of
states that is described by
\be\label{bulk}
\H_{\rm sphere} = \left( \H^{\widehat{\rm su}(2)}_0
\otimes {\bar\H^{\widehat{\rm su}(2)}_0} \right) \; \oplus \; \left(
\H^{\widehat{\rm su}(2)}_{\half} \otimes
{\bar\H^{\widehat{\rm su}(2)}_{\half}}\right) \,.
\ee
Because of the Sugawara construction, every highest weight
representation of $\su(2)_1$ also forms a representation of the
Virasoro algebra \Vir\ with $c=1$. The generators of the Virasoro
algebra commute with the current zero modes, and thus the
$\su(2)_1$ representations can be decomposed into representations
of su(2)$\;\oplus\;\Vir$. If we denote the $(2j+1)$-dimensional spin
$j$ representation of su(2) by $V^j$ and the Virasoro algebra
irreducible highest weight representation of weight $h$ by 
$\H^{\rm Vir}_h$, then we have 
\be\label{decompa}
\H^{\widehat{\rm su}(2)}_j
 = \sum_{n =0}^\infty  V^{(n+j)} \otimes \H^{\rm Vir}_{(n+j)^2} \,.
\ee
It thus follows that the space of states can be decomposed with
respect to the algebra 
su(2)$_L \oplus {\rm su(2)}_R \oplus \Vir_L \oplus \Vir_R$ as 
\be\label{decompb}
\H_{\rm sphere}
= \sum_{{j,\bar\jmath \in\Nop_0/2}\atop{j+\bar\jmath\in\Nop_0}}
  \left(
    V^{j} \otimes \bar V^{\bar\jmath}\right)
  \otimes
  \left(
   \H^{\rm Vir}_{j^2} \otimes \bar\H^{\rm Vir}_{\bar \jmath^2}
  \right)
\,.
\ee
As we have explained before, we can construct a (conformal) Ishibashi
state (\ie\ a state that satisfies (\ref{conformal}) but not
necessarily any additional gluing condition) for each tensor product 
$\H^{\rm Vir}_{h} \otimes \bar\H^{\rm Vir}_{\bar h}$ for which the two
Virasoro representations are conjugate, \ie\ for which
$h=\bar{h}$. Thus only the sectors with $j=\bar\jmath$ in the above 
decomposition give rise to Ishibashi states, and we obtain 
$V^{j} \otimes V^{j}$ conformal Ishibashi states for each
$j\in\half\Nop_0$. Let us denote these Ishibashi states by 
$\vert \, j;m,n\, \rangle\!\rangle$, where $m$ and $n$ run from
$-j,-j+1,\ldots,j-1,j$ and label the states in $V^j$. These Ishibashi
states are thus labelled just like matrix elements of su(2)
representations.  

The most general conformal boundary conditions have not been known
until recently \cite{GRW} (we shall give the complete solution below),
but a large class of su(2) preserving boundary states have been known
for some time \cite{GrGut1,RS1}. They are believed to be
consistent and fundamental since they can be obtained from the Cardy 
boundary states (that were shown to be consistent in
\cite{FFFS1,FFFS2}) by marginal deformations. These boundary states
are characterised by the gluing condition 
\be\label{affine}
\left( 
   {\rm Ad}_{(\,g \cdot\iota\,)}(J^a_m)  + \bar J^a_{-m} 
     \right) |\!|g\rangle\!\rangle  = 0 
\qquad {\rm where}\quad\  
  \iota = \pmatrix{0 & 1 \cr -1 & 0}\,.
\ee
In terms of the conformal Ishibashi states they are given as 
\be\label{su2brane}
|\!|g\rangle\!\rangle = {1\over 2^{1\over 4}}
\sum_{j,m,n} D^j_{m,n}(g)\, \vert \, j;m,n\, \rangle\!\rangle\,,
\ee
where $D^j_{m,n}(g)$ is the matrix element of $g$ in the
representation $j$, and $\vert \, j;m,n\, \rangle\!\rangle$
denotes the Virasoro Ishibashi state labelled by the triple $(j;m,n)$
as above. These boundary states include the Neumann and Dirichlet
branes we have discussed before; in fact, the boundary state
(\ref{Neubrane}) agrees with (\ref{su2brane}) for 
\be\label{Neugroup}
g = \left( \begin{array}{cc} 0 & e^{i{w\over 2\sqrt{2}}} \\ 
      - e^{-i{w\over 2\sqrt{2}}} & 0 \end{array} \right)\,,
\ee
while the Dirichlet brane (\ref{Dirbrane}) agrees with
(\ref{su2brane}) for
\be\label{Dirgroup}
g = \left( \begin{array}{cc} e^{i{a\over \sqrt{2}}} & 0 \\ 
   0 & e^{-i{a\over \sqrt{2}}} \end{array} \right)\,.
\ee
Provided that these boundary states are indeed consistent they have
to satisfy the factorisation constraint, and we therefore have to have
that 
\be\label{factor}
D^{j_1}_{m_1,n_1}(g)\, D^{j_2}_{m_2,n_2}(g) = \sum_{j;m,n}
M^{j;m,n}_{j_1; m_1,n_1 \, j_2;m_2,n_2} 
D^j_{m,n}(g)\,,
\ee
where we have denoted the (rescaled) structure constant 
$2^{1\over 4} C\cdot F$ from (\ref{ca}) summarily by $M$. This
equation has to hold for all  $g\in{\rm SU(2)}$ (since we have a
fundamental boundary condition for each such $g$). On the other hand,
$M$ does not depend on $g$ since it is given in terms of the operator
product coefficient $C$ and the fusing matrix $F$ that do not depend
on any boundary conditions. Thus we can use the {\it Peter-Weyl
Theorem} (see for example \cite{knapp})  to deduce that this can only
be the case if   
\be\label{Mresult}
M^{j;m,n}_{j_1; m_1,n_1 \, j_2;m_2,n_2} = 
(j_1m_1,j_2m_2|jm)\, (jn|j_1n_1,j_2n_2)\,,
\ee
where $(j_1m_1,j_2m_2|jm)$ and $(jn|j_1n_1,j_2n_2)$ are the
Clebsch-Gordan coefficients that describe the decomposition of the
tensor product  $j_1\otimes j_2$ in terms of the representation $j$.  
Indeed, the left hand side of (\ref{factor}) is the matrix element
between the states  labelled by $(m_1\otimes m_2)$ and 
$(n_1\otimes n_2)$ in the tensor product of the representations $j_1$
and $j_2$; the Clebsch-Gordan coefficients describe the decomposition
of this tensor product into irreducible representations, and therefore
these matrix elements must agree with the right hand side of
(\ref{factor}). Since the structure constants $M$ are uniquely
determined by (\ref{factor}), we may conclude that $M$ must be given
by (\ref{Mresult}).   

Thus we have succeeded to determine the structure constants of the
`classifying algebra' (\ref{ca}) from the knowledge of a sufficiently
large class of solutions. Given these structure constants we can now  
determine its most general solution; it is given by 
\be
B^{(j;m,n)} = {1\over 2^{1\over 4}} D^{j}_{m,n}(g) \qquad 
\hbox{where $g\in {\rm SL}(2,\Cop)$.} 
\ee
Thus we can conclude that the most general fundamental conformal
D-branes for this theory are (at most) labelled by group elements in 
${\rm SL}(2,\Cop)$ \cite{GRW}.

It is not difficult to check that all of these D-branes actually
satisfy the Cardy condition (see \cite{GRW} for a careful discussion
of this issue). The D-branes that are associated to  
SL$(2,\Cop)\backslash{\rm SU}(2)$ have complex couplings to
some of the space-time fields, and may therefore be unphysical (at
least from a string theory point of view). Apart from these unphysical
branes, all conformal D-branes actually preserve the full SU(2)
symmetry (since they satisfy the gluing condition
(\ref{affine})). This is quite surprising since the Virasoro condition
is (on the face of it) much weaker, and one would have expected to
find many more solutions.

\subsection{The conformal branes for other radii}

For $R\ne {1\over \sqrt{2}}$ we can use similar techniques to
determine all conformal D-branes of the theory, although the results
are not as complete as for the self-dual case
$R={1\over\sqrt{2}}$. The results depend on whether the radius of the
circle is a ratio of the self-dual radius, or whether it  is an
irrational multiple of the self-dual radius. In the former case, \ie\
if the radius is of the form 
\be
R= {M \over N} {1\over \sqrt{2}} \,,
\ee
where $M$ and $N$ are coprime positive integers, the most general
fundamental conformal D-branes can be described as follows \cite{GR}: 
every fundamental conformal D-brane is (i) either a Neumann or
Dirichlet brane (\ie\ has a boundary state given by (\ref{Neubrane})
or (\ref{Dirbrane}), respectively); or (ii) it is a brane associated
to an element in 
\be\label{quotient}
{\rm SU(2)} / \Zop_M \times \Zop_N \,,
\ee
that can be constructed by a formula similar to (\ref{su2brane}). 
(Details of the construction can be found in \cite{GR}.) If we write
an arbitrary group element of SU(2) as 
\be\label{groupelem}
g = \left( \begin{array}{cc} a & b \\ -b^\ast & a^\ast \end{array}
\right) \qquad |a|^2+|b|^2 = 1 \,,
\ee
then the generator of $\Zop_N$ acts as $a\mapsto e^{2\pi i\over N}a$,
while the generator of $\Zop_M$ acts as $b\mapsto e^{2\pi i \over M}b$.
The branes associated to (\ref{quotient}) are fundamental provided
that $ab\ne 0$; on the other hand, for $a=0$ the brane associated to 
(\ref{groupelem}) is the superposition of $N$ Neumann branes
(\ref{Neubrane}) with evenly spaced Wilson lines, while for 
$b=0$, the brane described by (\ref{groupelem}) is the superposition
of $M$ equidistantly spaced Dirichlet branes (\ref{Dirbrane}). The
general D-branes in the family interpolate between these two extremal
configurations. In fact, the Dirichlet or Neumann brane configurations
merge into intermediate boundary states that can no longer be thought
of as  superpositions of fundamental branes. These intermediate branes
are themselves fundamental, and do not preserve the U(1) symmetry.   

The situation at an `irrational' radius, \ie\ if $\sqrt{2} R$ is
irrational, can be formally deduced from the above by taking
simultaneously $M,N\rightarrow\infty$. In this limit the branes
labelled by (\ref{quotient}) then only depend on the modulus of $a$
and $b$. Since $|a|^2+|b|^2=1$, there is therefore only one real
parameter that we can take to be given by $x=2|a|^2-1$ with 
$-1\leq x \leq 1$. In addition to the standard Neumann and Dirichlet
branes the theory therefore has only an interval of branes labelled by
$x$ \cite{Friedan,Janik}. This interval of branes interpolates
between a smeared Dirichlet brane (\ie\ the integral of Dirichlet
boundary states where we integrate over all possible positions on the
circle) and a smeared Neumann brane (\ie\ the integral of Neumann
boundary states where we integrate over all possible Wilson lines on
the dual circle).   

\subsection{Introducing fermions}

Up to now we have only discussed bosonic conformal field
theories. For (world-sheet) fermions a few additional complications
arise. First of all, we need an additional parameter $\eta=\pm 1$ that
labels the different spin structures of the world-sheet. This
parameter modifies the gluing condition of every world-sheet
fermion. For example, the gluing condition for the superpartner $G$ of 
the stress-energy tensor is then
\be\label{superconf}
\bigl(\, G_r + i \eta\;\overline{G}_{-r}\,\bigr)\,
|\!|B,\eta\,\rangle\!\rangle = 0\,.
\ee
In the case with supersymmetry it is natural to consider only those
boundary conditions that not only preserve the conformal symmetry
(\ref{conformal}) but are also superconformal, \ie\ satisfy 
(\ref{superconf}).\footnote{In this lecture we are restricting
ourselves to the case with $N=1$ world-sheet supersymmetry. For
higher supersymmetry, one should also preserve the additional
supercurrents.} Finally, we also need to make sure that the boundary
states actually lie in the physical closed string spectrum, \ie\ that
they are GSO-invariant. In the one-dimensional case, only one
GSO-projection is consistent 
\cite{DGH},
\be
P^0_{\rm GSO} = \half\Bigl(1+(-1)^{F+\overline{F}}\Bigr)\,,
\ee
where $F$ and $\overline{F}$ are the left- and right-moving fermion
number operators. This is to be contrasted with the more familiar
GSO-projection 
\be\label{suGSO}
P_{\rm GSO} = {1\over 4} \Bigl(1+(-1)^F\Bigr)
\Bigl(1\pm(-1)^{\overline{F}}\Bigr)\,,
\ee
that is relevant for the ten-dimensional superstring theories. The
one-dimensional theory  we are considering here is therefore the
analogue of the type 0B (or 0A)  theory, rather than that of the type
IIB (or IIA) theory \cite{DixHar,seiwitt}.  

The analogue of the D-branes that preserve the full U(1) symmetry,
\ie\ that satisfy (\ref{Neuglu}) or (\ref{Dirglu}), satisfy now in
addition gluing conditions for the superpartner $\psi$ of the
current,  
\begin{eqnarray}
\left( \psi_r + i \eta \bar\psi_{-r} \right)\, 
|{\rm N},\eta\rangle\!\rangle 
& = & 0 \label{Neusu} \\
\left( \psi_r - i \eta \bar\psi_{-r} \right)\, 
|{\rm D},\eta\rangle\!\rangle 
& = & 0 \label{Dirsu} \,.
\end{eqnarray}
The most general such D-branes are conventional Neumann and Dirichlet
branes that are then either the analogue of BPS branes (as in
\cite{BG1}) or the analogue of the non-BPS branes of Sen \cite{Sen6}
(see also \cite{Thompson}).  The construction of these D-branes will
be briefly reviewed in the next section.

In addition to these U(1)-preserving D-branes there are also 
{\it superconformal} D-branes, \ie\ D-branes that only preserve
(\ref{conformal}) and (\ref{superconf}) but not necessarily 
(\ref{Neuglu}) and (\ref{Neusu}) or (\ref{Dirglu}) and (\ref{Dirsu}). 
If the radius of the circle is rational, $R={M\over N}$, where $M$ and
$N$ are coprime positive integers, then these additional D-branes are
associated to 
\be\label{quotientsu}
{\rm SU(2)} / \Zop_{\tilde{M}} \times \Zop_N \,,
\ee
where $\tilde{M}=2M$ if $N$ is odd, and $\tilde{M}=M$ otherwise. These
branes interpolate between brane anti-brane pairs and non-BPS branes
\cite{GR} (see Figure~1). This is consistent with charge conservation
since the overall R-R charge of both configurations vanishes. 
\pagebreak

\begin{figure}[h]
\vspace*{1cm}\hspace*{0.5cm}
\epsfxsize=14cm
\epsfbox{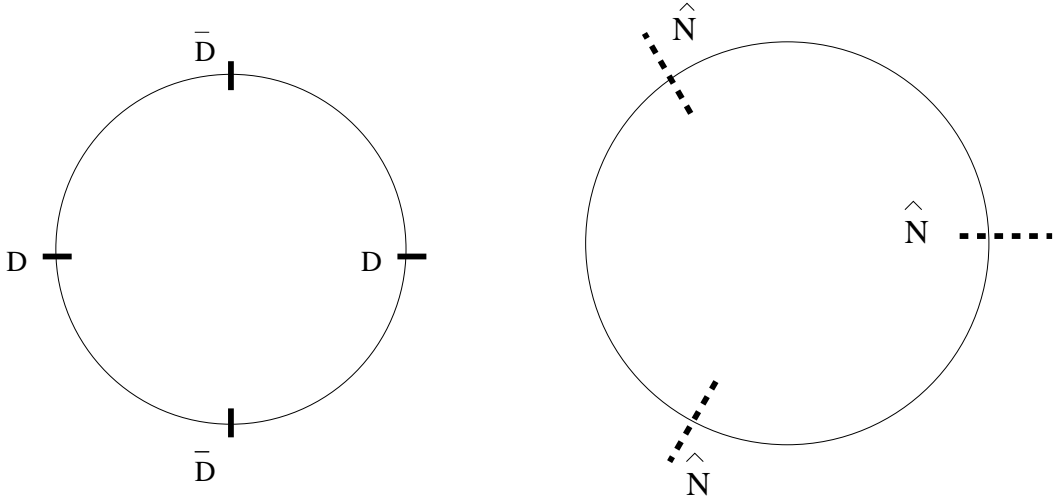}
\caption{The brane configurations for $R=2/3$ corresponding 
to special group elements: on the left, $g$ is of the form
(\ref{Dirgroup}) and the boundary state describes two Dirichlet branes
$D$ and two anti-Dirichlet branes $\overline{D}$ distributed evenly
over the target circle; on the right, $g$ is of the form
(\ref{Neugroup}), and we have three non-BPS Neumann branes  
$\hat N$ with Wilson lines that are evenly distributed over 
the dual circle.}
\end{figure}

\section{The ten-dimensional theories}

For the ten-dimensional string theories (that are ultimately of
interest) the complete set of superconformal D-branes is not yet
known. However, the branes that preserve the ten different U(1)s can
be easily described, and the restriction on the number of Neumann and 
Dirichlet directions for IIA/0A and IIB/0B can be straightforwardly
understood. The relevant construction was reviewed in detail in 
\cite{Gabrev}; we shall therefore be rather sketchy here, and only 
highlight some of the crucial features.

In the following we shall always work in light-cone gauge, and we
shall only consider D-branes for which the boundary condition for both 
light-cone directions is Dirichlet. We are interested in constructing
D-branes $|\!|{\rm Dp}\rangle\!\rangle$ for which $p+1$ of the ten
directions satisfy a Neumann boundary conditions. Since we are working
in light-cone gauge $p=-1,\ldots, 7$; the general case can be
obtained from this analysis by a double Wick rotation \cite{GrGut2}. 

Let us concentrate on the supersymmetric case in the following for
which the GSO projection is (\ref{suGSO}). In this case the
GSO-invariant D-branes are invariant under both $(-1)^F$ and 
$(-1)^{\overline{F}}$. Since
\be
(-1)^{\overline{F}}\, \left(\psi_r + i \eta \bar\psi_{-r} \right) 
= \left(\psi_r - i \eta \bar\psi_{-r} \right)\,(-1)^{\overline{F}}
\ee
the GSO-invariant D-branes are necessarily superpositions of boundary
states with both $\eta=\pm 1$. This fact is ultimately responsible for
the emergence of fermions in the open string spectrum.

In the NS-NS sector, every non-trivial boundary state (for fixed $\eta$)
can be made GSO-invariant (by adding to it a suitable boundary state
with the opposite sign for $\eta$), but in the R-R sector this is
not always possible. Indeed, in the R-R sector the boundary state must
satisfy a gluing condition coming from the fermionic zero modes (\ie\
from (\ref{Neusu}) and (\ref{Dirsu}) with $r=0$), and this is not
always compatible with the GSO-projection. As is explained in
\cite{Gabrev}, the boundary state 
$|\!|{\rm Dp}\rangle\!\rangle_{\rm R-R}$ is GSO-invariant provided
\be\label{IIAIIBcon}
\hbox{$p$ \hspace*{0.1cm} is} \quad \left\{\begin{array}{ll}
\hbox{even} \hspace*{1cm} & \hbox{for IIA} \\
\hbox{odd} \hspace*{1cm} & \hbox{for IIB.}
\end{array}\right.
\ee
As we have mentioned before, there are two types of U(1)-preserving
branes that are known. 

\subsection{Stable BPS branes}

The boundary states of the stable BPS D-branes are schematically
given as 
\be\label{BPS}
|\!|{\rm Dp}\rangle\!\rangle = |\!|{\rm Dp}\rangle\!\rangle_{\rm NS-NS}
\pm i |\!|{\rm Dp}\rangle\!\rangle_{\rm R-R} \,,
\ee
where $|\!|{\rm Dp}\rangle\!\rangle_{\rm NS-NS}$ and
$|\!|{\rm Dp}\rangle\!\rangle_{\rm R-R}$ are suitably normalised
superpositions of Ishibashi states in the NS-NS and the R-R sector,
respectively. The sign between the NS-NS and the R-R component in
(\ref{BPS}) distinguishes between branes and anti-branes. These
boundary states involve a R-R component, and they are therefore only
GSO-invariant provided that $p$ satisfies (\ref{IIAIIBcon}). The
D-branes satisfy Cardy's condition in that they lead to open strings
of the form 
\be
\hbox{[NS - R]}\; \half \Bigl(1+(-1)^F\Bigr) \,,
\ee
\ie\ to a total open string amplitude that is the trace over the
NS-sector minus the trace over the R-sector, both with the insertion
of the GSO-projector.\footnote{Since the states in the open string
R-sector correspond to spacetime fermions, they must contribute with a 
minus sign to this amplitude.} Because of this GSO-projection, the
tachyon from the open string NS-sector is removed, and the branes are
stable.

\subsection{Unstable non-BPS branes} 

The theory also possesses unstable {\it non-BPS} D-branes \cite{Sen6}
whose boundary states are of the form \cite{Horava}
\be
|\!|{\rm Dp}\rangle\!\rangle = \sqrt{2} \,
|\!|{\rm Dp}\rangle\!\rangle_{\rm NS-NS}\,.
\ee
Here $|\!|{\rm Dp}\rangle\!\rangle_{\rm NS-NS}$ is the same
combination of Ishibashi states (including normalisation) that
appeared in (\ref{BPS}). As is explained in detail in \cite{Gabrev},
these branes occur for the complementary values of $p$ relative to 
(\ref{IIAIIBcon}). Their open string spectrum is now of the form 
\be
\hbox{[NS - R]}
\ee
without a GSO-projection. As a consequence the open string tachyon
from the NS-sector is not removed, and the D-brane is unstable.

\subsection{Stable non-BPS branes}

As we have seen above, all stable (fundamental) U(1)-preserving
D-branes of Type IIA/IIB string theory in flat space are actually
BPS. However, this is not true in general. In particular, stable
non-BPS D-branes exist for certain orbifold theories of Type
IIA/IIB. The simplest example is the D0-brane of the orbifold of IIB 
on $T^4/(-1)^{F_L}I_4$ \cite{BG2,Sen2}. Here $I_4$ denotes the
inversion of the four torus directions, and $(-1)^{F_L}$ acts as  
$\pm 1$ on left-moving space-time fermions. The corresponding boundary
state is schematically of the form
\be
|\!|{\rm D0}\rangle\!\rangle = |\!|{\rm D0}\rangle\!\rangle_{\rm NS-NS}
\pm i |\!|{\rm D0}\rangle\!\rangle_{\rm R-R;T} \,,
\ee
where the last component is in the twisted R-R sector. This boundary
state is not BPS since it does not have a component in the untwisted
R-R sector. However, it is nevertheless stable (at least for
sufficiently large radii) since the corresponding open string is
\be
\hbox{[NS - R]}\; \half \Bigl(1+(-1)^F I_4\Bigr) \,.
\ee
This projection removes the zero-winding component of the tachyon in
the open string NS-sector, and thus stabilizes the D-brane.  

Stable non-BPS D-branes play an important role for understanding
string dualities of supersymmetric string theories. For example the
T-dual of the above D0-brane is a non-BPS D1-brane for IIA on
$T^4/I_4$ \cite{Sen6,BG3}. Since $T^4/I_4$ is an orbifold limit of K3,
this theory is dual to the Heterotic string on $T^4$. The dual of the
non-BPS D1-brane of the IIA theory can then be identified with a
certain perturbative stable non-BPS state of the Heterotic theory
\cite{BG3}. This sheds some light on how string duality relates states
that are not BPS, and that are therefore not protected from quantum
corrections.

\section{Conclusions} 

In this lecture I have described some of the key features in the
construction of D-branes using conformal field theory techniques. As
should have become apparent, this is a powerful construction that
gives rise to exact string theoretic results. 

I have described in detail the complete classification of the
conformal D-branes for the simple case of a single free boson, and the
supersymmetric analogue of a single boson and a single Majorana
fermion. I have also sketched the more familiar construction of the
standard U(1)-preserving D-branes for a number of ten-dimensional
theories.  

There are a number of avenues that it would be interesting to
explore. In particular, it would be important to understand all
superconformal D-branes for the ten-dimensional string theories. It
would also be interesting to classify all the D-branes that preserve
$N=2$ world-sheet supersymmetry. Indeed, most of the additional
superconformal D-branes we have constructed are non-BPS and in fact
unstable. In order to obtain space-time supersymmetric (stable)
D-branes, one presumably needs to preserve $N=2$ supersymmetry on the
world-sheet. Thus it would be interesting to see whether our 
analysis can be extended to that case.

\section*{Acknowledgements}

I would like to thank the organisers for organising a very stimulating
and interesting school, and for giving me the opportunity to present
these lectures. I thank Andreas Recknagel and G\'erard Watts for
collaborations and useful discussions on many issues discussed in
these lectures. I am also grateful to Hanno Klemm and Andreas
Recknagel for a careful reading of a draft of these notes. 

I am grateful to the Royal Society for a University Research
Fellowship. This work is also partly supported by EU contract
HPRN-CT-2000-00122, and the  PPARC special grant ``String Theory and
Realistic Field Theory'', PPA/G/S/1998/0061.



\begin{thebibliography}{77}



\bibitem{MM} G. Moore and R. Minasian, J. High Energy Phys. 
{\bf 9711} (1997) 002 [hep-th/9710230].

\bibitem{WittenK} E. Witten, J. High Energy Phys. {\bf 9812} (1998)
019 [hep-th/9810188]. 

\bibitem{Douglas} M. Douglas, {\it D-branes, categories and $N=1$
supersymmetry} [hep-th/0011017].

\bibitem{dm} M.R. Douglas and G. Moore, {\it D-branes, quivers, and
ALE instantons} [hep-th/ 9603167].

\bibitem{Frau1} M. Frau, I. Pesando, S. Sciuto, A. Lerda and R. Russo,
Phys. Lett. {\bf B400} (1997) 52 [hep-th/9702037].

\bibitem{Frau2} P. Di Vecchia, M. Frau, I. Pesando, S. Sciuto,
A. Lerda and R. Russo, Nucl. Phys. {\bf B507} (1997) 259
[hep-th/9707068]. 

\bibitem{Sen6} A. Sen, J. High Energy Phys. {\bf 9812} (1998) 021
[hep-th/9812031]. 

\bibitem{BG3} O. Bergman and M.R. Gaberdiel, J. High Energy Phys.
{\bf 9903} (1999) 013 [hep-th/9901014]. 

\bibitem{DDG} D-E. Diaconescu, M.R. Douglas and J. Gomis, 
J. High Energy Phys. {\bf 9802} (1998) 013 [hep-th/9712230].  

\bibitem{DG} D-E. Diaconescu and J. Gomis, J. High Energy Phys.
{\bf 0010} (2000) 001 [hep-th/9906242]. 

\bibitem{GabSen} M.R. Gaberdiel and A. Sen, J. High Energy Phys.
{\bf 9911} (1999) 008 [hep-th/9908060].

\bibitem{GS} M.R. Gaberdiel and B. Stefa\'nski, Nucl. Phys. {\bf B578}
(2000) 58 [hep-th/9910109].

\bibitem{RS} A. Recknagel and V. Schomerus, Nucl. Phys. {\bf B531}
(1998) 185 [hep-th/9712186].  

\bibitem{GutSat} M. Gutperle and Y. Satoh, Nucl. Phys. {\bf B555}
(1999) 477 [hep-th/9902120].

\bibitem{FSW} J. Fuchs, C. Schweigert and J. Walcher, Nucl. Phys.
{\bf B588} (2000) 110 [hep-th/0003298].

\bibitem{BDLR} I. Brunner, M.R. Douglas, A. Lawrence and
Ch. R\"omelsberger, J. High Energy Phys. {\bf 0008} (2000) 015
[hep-th/9906200].

\bibitem{GRW} M.R. Gaberdiel, A. Recknagel and G.M.T. Watts, {\it The
conformal boundary states for $SU(2)$ at level 1} [hep-th/0108102].  

\bibitem{GR} M.R. Gaberdiel and A. Recknagel, J. High Energy Phys.
{\bf 0111} (2001) 016 [hep-th/0108238].

\bibitem{Vafa} C. Vafa, Phys. Lett. {\bf B199} (1987) 195.

\bibitem{Sonoda} H. Sonoda, Nucl. Phys. {\bf B311} (1989) 401.

\bibitem{MooreSei} G. Moore and N. Seiberg, Commun. Math. Phys.
{\bf 123} (1989) 177.

\bibitem{lew} D.C. Lewellen, Nucl. Phys. {\bf B372} (1992) 654.

\bibitem{carlew} J.L. Cardy and D.C. Lewellen, Phys. Lett. 
{\bf B259} (1991) 274.

\bibitem{PSS1} G. Pradisi, A. Sagnotti and Ya. S. Stanev,
Phys. Lett. {\bf B381} (1996) 97 [hep-th/9603097].

\bibitem{Ingo1} I. Runkel, Nucl. Phys. {\bf B549} (1999) 563
[hep-th/9811178]. 

\bibitem{BPPZ} R.E. Behrend, P.A. Pearce, V.B. Petkova and
J.-B. Zuber, Nucl. Phys. {\bf B579} (2000) 707 [hep-th/9908036].

\bibitem{Ingo2} I. Runkel, Nucl. Phys. {\bf B579} (2000) 561
[hep-th/9908046].

\bibitem{FFFS1} G. Felder, J. Fr\"ohlich, J. Fuchs and C. Schweigert,
Phys. Rev. Lett. {\bf 84} (2000) 1659 [hep-th/9909140].

\bibitem{FFFS2} G. Felder, J. Fr\"ohlich, J. Fuchs and C. Schweigert,
{\it Correlation functions and boundary conditions in RCFT and
three-dimensional topology} [hep-th/9912239].

\bibitem{Gannon} T. Gannon, {\it Boundary conformal field theory and
fusion ring representations} [hep-th/0106105].

\bibitem{Ishibashi} N. Ishibashi, Mod. Phys. Lett. {\bf A4} (1989)
251.  

\bibitem{Cardy} J.L. Cardy, Nucl. Phys. {\bf B324} (1989) 581.

\bibitem{Verlinde} E. Verlinde, Nucl. Phys. {\bf B300} (1988) 360. 

\bibitem{RS1} A. Recknagel and V. Schomerus, Nucl. Phys. {\bf B545}
(1999) 233 [hep-th/9811237].

\bibitem{FS} J. Fuchs and C. Schweigert, Phys. Lett. {\bf B414} (1997)
251 (1997) [hep-th/9708141].

\bibitem{GrGut1} M.B. Green and M. Gutperle, Nucl. Phys. {\bf B460}
(1996) 77 [hep-th/9509171].

\bibitem{knapp} A.W. Knapp, {\it Representation theory of semisimple 
groups: an overview based on examples}, Princeton University Press,
Princeton (1986).

\bibitem{Friedan} D. Friedan, {\it The space of conformal boundary
conditions for the $c=1$ Gaussian model}, unpublished note (1999). 

\bibitem{Janik} R.A. Janik, Nucl. Phys. {\bf B618} (2001) 675
[hep-th/0109021].

\bibitem{DGH} L.J. Dixon, P. Ginsparg and J.A. Harvey, Nucl. Phys. 
{\bf B306} (1988) 470. 

\bibitem{DixHar} L.J. Dixon and J.A. Harvey, Nucl. Phys. {\bf B274}
(1986) 93.

\bibitem{seiwitt} N. Seiberg and E. Witten, Nucl. Phys. {\bf B276}
(1986) 272.

\bibitem{BG1} O. Bergman and M.R. Gaberdiel, Nucl. Phys. {\bf B499}
(1997) 183 [hep-th/9701137].

\bibitem{Thompson} D.M. Thompson, {\it Descent relations in type 0A/0B}
[hep-th/0105314].

\bibitem{Gabrev} M.R. Gaberdiel,  Class. Quant. Grav. {\bf 17} (2000)
3483 [hep-th/0005029]. 

\bibitem{GrGut2} M.B. Green and M. Gutperle, Nucl. Phys. {\bf B476}
(1996) 484 [hep-th/9604091].

\bibitem{Horava} P. Ho\v{r}ava, Adv. Theor. Math. Phys. {\bf 2} (1998)
1373 [hep-th/9812135].

\bibitem{BG2} O. Bergman and M.R. Gaberdiel, Phys. Lett. {\bf B441}
(1998) 133 [hep-th/9806155].

\bibitem{Sen2} A. Sen, J. High Energy Phys. {\bf 9808} (1998) 010
[hep-th/9805019]. 


\end{thebibliography}
\end{document}